\documentclass[aps,prd,preprint,groupedaddress,showpacs]{revtex4-1}
\usepackage{amssymb}
\usepackage{graphicx}
\usepackage{dcolumn}
\usepackage{bm} 

\newcommand {\func}[1]{\,\mathrm{#1}\,}

\begin{document}

\title{Passage of test particles through oscillating spherically-symmetric dark matter configurations}
\author{Vladimir A. Koutvitsky}
\author{Eugene M. Maslov}
\email{zheka@izmiran.ru}
\affiliation{Pushkov Institute of Terrestrial Magnetism, Ionosphere and Radio Wave Propagation (IZMIRAN) 
of the Russian Academy of Sciences,\\ Moscow, Troitsk, Kaluzhskoe Hwy 4, Russian Federation, 108840}
\date{\today}

\begin{abstract}
Applying the perturbative approach to geodesic equations, we study motion of
the test particles in time-dependent spherically symmetric spacetimes created by
oscillating dark matter. Assuming the weakness of the gravitational field,
we derive general formulas that describe infinite trajectories of the test
particles and determine the total deflection angle in the leading order
approximation. The obtained formulas are valid for both time-dependent and
static matter configurations. Using these results, we calculate the
deflection angle of a test particle passing through a spherically symmetric
oscillating distribution of a self-gravitating scalar field with a
logarithmic potential. It turned out that, in a wide range of amplitudes,
oscillations in the deflection angle are sinusoidal and become small 
for ultrarelativistic particles.
\end{abstract}

\maketitle

\section{Introduction}

Despite the great efforts of theorists and significant advances in
technologies and methods of observation, the nature of dark matter (DM)
remains unknown. The standard $\Lambda $CDM model, which explains
observational data well at cosmological scales, faces serious problems at
galactic and subgalactic scales (see, e.g., \cite{Primack}). One of the most
cited possibilities to overcome these difficulties is to assume that DM
consists of ultra-light bosonic particles with masses in the $%
10^{-23}-10^{-21}$ eV range, e.g., axions \cite{Marsh}, which are in a
coherent state described by a classical scalar field \cite{Turner, Seidel1,
Seidel2, Kolb, Lee, Peebles, Hu, Matos, Magana, Hui}. In the early Universe, 
the primordial fluctuations of this field are stretched by inflation, 
this evolution resulting in the formation of a uniform scalar background 
oscillating near the minimum of the effective potential.
These oscillations are unstable \cite%
{Khlopov}. In the case of the quadratic potential, the oscillating
background behaves as a dust-like matter, so that some kind of the Jeans
instability can occur \cite{Hu}. In addition, in the case of a
self-interacting scalar field, another instability mechanism comes into
play. This mechanism is based on parametric resonance between the
oscillating background and perturbations and works on both cosmological and
astrophysical scales \cite{Kofman, Koutvitsky0, Amin, Zhang, Lozanov,
Koutvitsky1, Fukunaga, Arvanitaki}. At the nonlinear stage this leads to
formation of quasi-stable oscillating lumps, oscillons (pulsons), (see \cite%
{Olle} for a recent review). 
Under the influence of gravity, after the completion of some relaxation processes, 
these lumps turn into long-lived self-gravitating oscillating objects, 
oscillatons (gravipulsons) \cite{Urena, Fodor, Koutvitsky2}, separated from the Hubble flow. 
The latter
means that the dynamics of an individual oscillaton should be determined by
the self-consistent system of Einstein-Klein-Gordon equations. Note that
oscillatons can arise from rather arbitrary localized initial conditions due
to the gravitational cooling process \cite{Seidel1, Seidel2, Guzman}. This
process is very similar to that which occurs in the integrable systems, when
an initial state decays into solitons and outgoing waves.

On galactic and subgalactic scales, oscillaton solutions can describe
various localized objects, from oscillating soliton stars to oscillating DM
halos, depending on the assumed mass of the scalar field. Oscillations of
the scalar field in these objects cause oscillations of the gravitational
potential, which can be detected by their effect on the motion of photons
and test bodies. In particular, as shown in \cite{Khmelnitsky}, the
gravitational time delay for a photon passing through an oscillating halo
should cause small periodic fluctuations in the observed timing array of the
pulsar located inside the halo. Although the predicted effect is very small, 
the authors believe it can be detected in the next generation of pulsar timing observations.
In paper \cite{Aoki}, it was proposed to use the laser interferometers for detecting 
the axion wind caused by passage of the Earth through DM.
The gravitational field oscillations,
produced by the oscillating DM, look like gravitational waves to an observer
on Earth and would be detected in future laser interferometer experiments.
An approach based on the observations of binary pulsars was discussed in 
\cite{Blas,Rozner} to probe of ultralight axion DM. 
It was shown that oscillations of DM resonantly perturb  the orbits of the binary pulsars
thus leading to secular variations in their orbital period.
Also, in the context of
oscillating DM, in Refs. \cite{Becerril, Boskovic} the orbital motion of
test bodies in spherically symmetric time-periodic spacetimes was studied
numerically. In particular, it was demonstrated in \cite{Boskovic} that the
orbital resonances may occur in motion of stars in oscillating spherically
symmetric halos. In addition, in Refs. \cite{Boskovic, Koutvitsky3} it was
shown that spectroscopic emission lines from stars in such halos exhibit
characteristic, periodic modulation patterns due to variations in the
gravitational frequency shift. These results show that the motion of photons
and test bodies may carry distinguishable observational imprints of the
oscillating DM.

Recently, in the above context, we studied the deflection of photons in
time-periodic spherically symmetric gravitational fields \cite{Koutvitsky4}.
Using the geodesic method and the perturbative approach, we have shown that
the deflection angle of a light ray in general undergoes periodic variations
when passing through such fields. In observations, this can lead to
additional variations of intensity of images when lensing the distant
sources. In the present paper, following the approach developed in \cite%
{Koutvitsky4}, we study the deflection of massive particles.

Our paper is organized as follows. In Sec. II, assuming the weakness of the
gravitational field, we use the perturbative approach to describe the
infinite trajectories of massive particles in nonstatic spherically
symmetric spacetimes. In particular, we obtain general formulas which
determine the deflection angle of a massive particle in the leading order
approximation. In Sec. III, we apply these formulas to calculate the
deflection angle of a massive test particle passing through an oscillating
dark matter configuration formed by a real scalar field with a logarithmic
self-interaction. Discussion and concluding remarks can be found in
Sec. IV.

\section{Infinite trajectories of test particles in nonstatic\\ spherically
symmetric spacetimes}

Let us consider a spherically symmetric nonstatic metric of the form%
\begin{equation}
ds^{2}=B(t,r)\,dt^{2}-A(t,r)\,dr^{2}-r^{2}(d\vartheta ^{2}+\sin
^{2}\vartheta \,d\varphi ^{2}),  \label{eq1}
\end{equation}%
where $A(t,r)$ and $B(t,r)$ tend to unity as $r\rightarrow \infty $. For the
trajectories lying in the plane $\vartheta =\pi /2$, the geodesic equation
reduces to the system

\begin{equation}
\frac{d}{ds}\ln \left( B\frac{dt}{ds}\right) =\frac{\dot{B}}{2B}\frac{dt}{ds}%
-\frac{\dot{A}}{2B}\left( \frac{dr}{ds}\right) ^{2}\left( \frac{dt}{ds}%
\right) ^{-1},  \label{eq2}
\end{equation}%
\begin{equation}
\frac{d^{2}r}{ds^{2}}+\frac{B^{\prime }}{2A}\left( \frac{dt}{ds}\right) ^{2}+%
\frac{\dot{A}}{A}\frac{dt}{ds}\frac{dr}{ds}+\frac{A^{\prime }}{2A}\left( 
\frac{dr}{ds}\right) ^{2}-\frac{r}{A}\left( \frac{d\varphi }{ds}\right)
^{2}=0,  \label{eq3}
\end{equation}%
\begin{equation}
\frac{d^{2}\varphi }{ds^{2}}+\frac{2}{r}\frac{dr}{ds}\frac{d\varphi }{ds}=0,
\label{eq4}
\end{equation}%
where 
$(\dot{\phantom{.}})=\partial /\partial t$, $\left( ^{\prime }\right)
=\partial /\partial r$. From Eqs. (\ref{eq4}) and (\ref{eq1}) it follows
that%
\begin{equation}
\frac{d\varphi }{ds}=\frac{J}{r^{2}},  \label{eq5}
\end{equation}%
\begin{equation}
A\left( \frac{dr}{ds}\right) ^{2}-B\left( \frac{dt}{ds}\right) ^{2}+\frac{%
{J}^{2}}{r^{2}}+1=0,  \label{eq6}
\end{equation}%
where ${J}=const$. It is easy to see that for a particle (e.g., of unit mass)
coming from a distant point with an initial velocity $v$ and an impact
parameter $b$%
\begin{equation}
{J}=bv\,{E},\quad {E}=\left( 1-v^{2}\right) ^{-1/2},  \label{eq7}
\end{equation}%
so that ${J}$ is the particle's angular momentum, ${E}$ is the initial kinetic
energy.

Let us assume that the gravitational field is time-dependent and weak
everywhere on the particle trajectory, i.e.,%
\begin{equation}
A=1-2\psi +O(\varkappa ^{2}),\quad B=1+2\chi +O(\varkappa ^{2}),  \label{eq8}
\end{equation}%
where $\psi (t,r)$ and $\chi (t,r)$ are small functions of order $\varkappa $,
$\varkappa\ll 1$  being a dimensionless small parameter proportional to
the gravitational constant $G$.
\begin{figure}
\includegraphics[width=7cm]{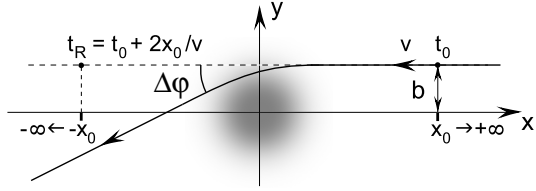}
\caption{Passage of the test particle through the gravitating mass}
\end{figure}

Now suppose that in the $xy$ plane at a distant point $x=x_{0}$, $y=b$ at a
moment $t_{0}$, a particle begins to move with an initial velocity $v$
parallel to the $x$ axis in the direction of the gravitating mass (see Fig.
1). If the gravitating mass were absent, the particle would move along the
straight line,%
\begin{equation}
x=x_{0}+v(t_{0}-t),\quad y=b,  \label{eq9}
\end{equation}%
and be registered at the moment $t_{R}=t_{0}+2x_{0}/v$ at the distant point $%
x=-x_{0}$, $y=b$. On this line%
\begin{equation}
r(t)=\sqrt{x^{2}(t)+b^{2}},\quad dt/ds=E.  \label{eq10}
\end{equation}

With the gravitating mass, the particle will move along a deflected
trajectory with the current radial coordinate%
\begin{equation}
r(t)=\left( 1+\eta (t)\right) \sqrt{x^{2}(t)+b^{2}},  \label{eq11}
\end{equation}%
where, as before,%
\begin{equation}
x=v(t_{R}-t)-x_{0},  \label{eq12}
\end{equation}%
and $\eta (t)$ is a small function of order $\varkappa $. On this
trajectory, the dependence $t(s)$ is determined by Eq. (\ref{eq2}), where we
set%
\begin{equation}
B\frac{dt}{ds}=E\left( 1+\zeta (t)\right) ,  \label{eq13}
\end{equation}%
with a small function $\zeta (t)\sim \varkappa $. Then, with the required
accuracy, from Eq. (\ref{eq2}) we obtain%
\begin{equation}
\frac{d\zeta }{dt}=\dot{\chi}(t,r)+v^{2}\dot{\psi}(t,r)\left( 1-\frac{b^{2}}{%
r^{2}}\right) .  \label{eq14}
\end{equation}

To get the equation for $\eta $, we proceed from Eq. (\ref{eq6}), where $%
dr/ds=(dr/dt)\left( dt/ds\right) $, and $r$ is now given by Eq. (\ref{eq11}).
Calculating $dr/dt$ and using Eqs. (\ref{eq7}), (\ref{eq8}), and (\ref{eq13}%
), in the first order in $\varkappa $ we arrive at the equation%
\begin{eqnarray}
&&vx(x^{2}+b^{2})\frac{d\eta }{dt}-v^{2}(x^{2}-b^{2})\eta +v^{2}x^{2}\psi(t,r)+\nonumber\\
&&\left[ \left( 2v^{2}\!-\!1\right) x^{2}\!-\!b^{2}\right]\! \chi (t,r)+
\left[\left( 1\!-\!v^{2}\right) x^{2}\!+\!b^{2}\right] \zeta (t)=0.\qquad  \label{eq15}
\end{eqnarray}

Further, from Eq. (\ref{eq5}), using Eqs. (\ref{eq7}), (\ref{eq8}), (\ref%
{eq11}), and (\ref{eq13}), we find

\begin{equation}
\frac{d\varphi }{dt}=\frac{vb}{x^{2}+b^{2}}\left[ 1+\left( 2\chi -\zeta
-2\eta \right) \right] .  \label{eq16}
\end{equation}

Since in Eqs. (\ref{eq14})-(\ref{eq16}) $t=t_{R}-\left( x+x_{0}\right) /v$
and $r=\sqrt{x^{2}+b^{2}}$, we can integrate over $x$ instead of $t$,
setting $dx=-v\,dt$. As a result, we obtain%

\begin{eqnarray}
&&\zeta =\frac{1}{v}\int_{x}^{x_{0}}\left[ \dot{\chi}(t,r)+v^{2}\dot{\psi}%
(t,r)\left( 1-\frac{b^{2}}{r^{2}}\right) \right] \;dx,  \label{eq17}\\
&&\eta =\frac{x}{v^{2}\left( x^{2}+b^{2}\right) }\times\nonumber\\
&&\left\{\! \int \bigg[v^{2}x^{2}\psi (t,r)+\left( \left( 2v^{2}\!-\!1\right) x^{2}-b^{2}\right) \chi(t,r) 
+\left( \left( 1\!-\!v^{2}\right) x^{2}+b^{2}\right) \zeta (t)\bigg] \frac{dx}{x^{2}} +const\right\} ,  \label{eq18}\\
&&\varphi =\pi /2-\func{arctg}(x/b)+b\int_{x}^{x_{0}}\frac{2\chi -\zeta -2\eta}{x^{2}+b^{2}}\,dx.  \label{eq19}
\end{eqnarray}%

These equations completely describe, in the leading order, the trajectory of
the test particle that was emitted at a distant point with the coordinates $%
(x_{0},b)$ and registered by a distant observer at the moment $t_{R}$. The $%
constant$ in Eq. (\ref{eq18}) can be found from the condition $\eta
(x_{0})=0 $, but it does not affect the complete change of $\varphi $ for
the particle coming from infinity and going to infinity. Indeed, taking $%
x=-x_{0}$ and setting $x_{0}\rightarrow \infty $, we find $\varphi =\pi
+\Delta \varphi $, where%
\begin{equation}
\Delta \varphi =b\int_{-\infty }^{\infty }\frac{2\chi -\zeta -2\eta }{%
x^{2}+b^{2}}\,dx  \label{eq20}
\end{equation}%
is the deflection angle.

The obtained formulas are valid not only for time-dependent metrics, but
also for the static ones. In the latter case one should put $\zeta =0$ in
accordance with Eq. (\ref{eq17}). Consider, for example, the Schwarzschild
metric. Assuming $r_{g}/b=\varkappa \ll 1$, where $r_{g}=2GM$ is the
gravitational radius, we have%
\begin{equation}
\psi =\chi =-\varkappa \frac{b}{2r}.  \label{eq21}
\end{equation}%
Then, formula (\ref{eq18}) gives%
\begin{equation}
\eta =-\varkappa\, \frac{bx}{2v^{2}(x^{2}+b^{2})}
\left[ \frac{\sqrt{x^{2}+b^{2}}}{x}+\left( 3v^{2}-1\right) \func{arsh}\frac{x}{b}+const\right]\!.
\label{eq22}
\end{equation}%
Substituting (\ref{eq21}) and (\ref{eq22}) into (\ref{eq20}) and
integrating, we reproduce the well-known result,%
\begin{equation}
\Delta \varphi =\frac{2GM}{bv^{2}}\left( 1+v^{2}\right) +O((r_{g}/b)^{2}).
\label{eq23}
\end{equation}%
Note that this formula is only valid for $v^{2}\gg \varkappa =2GM/b$. 

In the case of a time-dependent metric, the deflection angle for a large
fixed $x_{0}$ will generally depend on the particle emission time $t_{0}$
or, equivalently, on the observation time $t_{R}=t_{0}+2x_{0}/v$, since when
integrating in (\ref{eq17})-(\ref{eq20}) we substitute $t=t_{R}-\left(
x+x_{0}\right) /v$ into the potentials $\psi (t,r)$ and $\chi (t,r)$. For
time-periodic potentials (with a certain period $T_{g}$), we can ignore $%
x_{0}$ in the integrands by setting for convenience $x_{0}=nT_{g}v$, where $%
n $ is a large integer. Then the moment $t_{R}$ will determine in which
phase of the oscillations of the gravitational field the particle passed
through the matter distribution and, consequently, at what angle it
deflected as a result of this.

In the next section, we calculate the deflection angle of the test particle
passing through the oscillating distribution of a real scalar field with a
logarithmic self--interaction.

\section{Deflection of the test particle by a time-periodic\\ spherically
symmetric scalar field}

As a deflecting matter, we consider the self-gravitating real scalar field
with the potential%
\begin{equation}
U(\phi )=\frac{m^{2}}{2}\phi ^{2}\left( 1-\ln \frac{\phi ^{2}}{\sigma ^{2}}%
\right) ,  \label{eq24}
\end{equation}%
where $\sigma $ is the characteristic magnitude of the field, $m$ is the
mass (in units $\hbar =c=1)$. Originally, such potentials were considered in
quantum field theory \cite{Rosen, Birula}. Also, when taking into account
quantum corrections, they naturally appear in inflationary cosmology \cite%
{Linde, Barrow}, as well as in some supersymmetric extensions of the
Standard Model \cite{Enqvist}. It is remarkable that potential (\ref{eq24})
admits exact solutions of the Klein-Gordon equation in the form of
multidimensional localized time-periodic field configurations, the pulsons
(oscillons) \cite{Marques, Bogolubsky, Maslov}. The corresponding solution
of the Einstein--Klein--Gordon system was found in paper \cite{Koutvitsky2}
by the Krylov-Bogoliubov method. This solution describes a self-gravitating 
field lump of an almost Gaussian shape that pulsates in time.
In the weak field approximation, the corresponding metric functions $A(t,r)$ and $B(t,r)$ can
be written as (\ref{eq8}), where%
\begin{equation}
\psi (t,r)\!=\frac{\varkappa }{2}\!\left[ V_{\max }\!\left( 1\!-\!\frac{\sqrt{\pi }\,%
\mathrm{erf}\,\rho }{2\rho }e^{\rho ^{2}}\!\right) \!+\!a^{2}\rho ^{2}\right]\!
e^{3-\rho ^{2}},  \label{eq25}\\
\end{equation}
\vspace{-12pt}
\begin{equation}
\chi (t,r)\!=\!-\frac{\varkappa }{2}\!\left[ V_{\max }\!\left( 1\!+\!\frac{\sqrt{\pi }\,%
\mathrm{erf}\,\rho }{2\rho }e^{\rho ^{2}}\!\right) \!+\!a^{2}\ln a^{2}\right]\!
e^{3-\rho ^{2}},  \label{eq26}
\end{equation}
$\tau =mt$, $\rho =mr$, $\varkappa =4\pi G\sigma ^{2}\ll 1$ ($G$ is the
gravitational constant). The function $a(\theta (\tau ))$ oscillates in the
range $-a_{\max }\leqslant a(\theta )\leqslant a_{\max }$ in the local
minimum of the potential $V(a)$,%
\begin{eqnarray}
a_{\theta \theta }&=&-dV/da, \label{eq27} \\ 
V(a)&=&(a^{2}/2)\left( 1-\ln a^{2}\right),  \label{eq28}
\end{eqnarray}
where $V_{\max }=V(a_{\max })$, $\theta _{\tau }=1+\varkappa \Omega
+O(\varkappa ^{2})$, and the constant $\varkappa \Omega $ is the frequency
correction due to gravitational effects (see Ref. \cite{Koutvitsky2} for
details). The period (in $\theta $) of these oscillations is given by%
\begin{equation}
T=4\int_{0}^{1}\left[ (1-\ln a_{\max }^{2})(1\!-\!z^{2})+z^{2}\ln z^{2}\right]
^{-1/2}dz.  \label{eq29}
\end{equation}%
The dependence of the period on $a_{\max }^{2}$ is shown in Fig. 2. With $%
a_{\max }^{2}\ll 1$ it can be approximated by 
$T\approx 2\pi \left( 1-\ln a_{\max }^{2}\right)^{-1/2}$. 

\begin{figure}
\includegraphics[width=8.5cm]{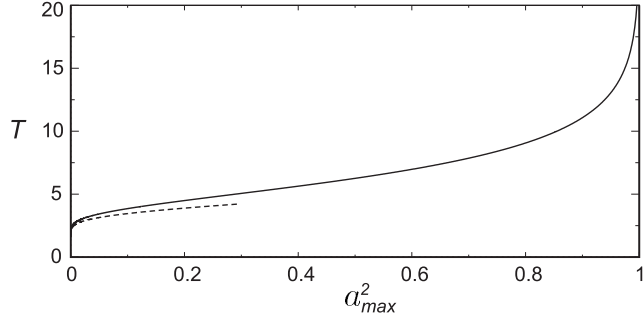}
\caption{Field oscillation period versus $a^2_{max}$ (solid line) and its approximation 
for small amplitudes  (dashed line)}
\end{figure}

The energy density of the field lump we are considering is concentrated on
the characteristic scale $r\sim m^{-1}$. As seen from Eqs. (\ref{eq25}) and (%
\ref{eq26}), at large distances from the lump the gravitational field turns
into the static Schwarzschild field (\ref{eq21}) with the mass $M=\left( e\sqrt{\pi }\right) ^{3}\sigma ^{2}m^{-1}V_{\max }$, 
in accordance with the
Birkhoff theorem (see, e.g., \cite{Weinberg}). However, inside the lump the
gravitational field oscillates with the period $T_{g}=[2m(1+\varkappa \Omega
)]^{-1}T$ (with respect to $t$).

Let us calculate the effect of these oscillations on the deflection angle of
the test particle passing through the lump. First of all, we need to find
the functions $\zeta $ and $\eta $ by formulas (\ref{eq17}) and (\ref{eq18}%
). Calculating $\dot{\psi}(t,r)$, $\dot{\chi}(t,r)$ and setting
\begin{eqnarray*}
{}&&\tau =\tau _{R}-(\xi +\xi _{0})/v,\quad \rho =\sqrt{\xi ^{2}+\beta^{2}},\quad \xi =mx,\\
{}&&\beta =mb,\quad \xi _{0}=mx_{0},\quad \tau _{R}=mt_{R},\quad d/d\tau=-v\,d/d\xi ,
\end{eqnarray*}
from (\ref{eq17}) we find%
\begin{equation}
\zeta =\frac{\varkappa }{2}e^{3-\beta ^{2}}\!\!\int_{\xi }^{\xi _{0}}\!\left[ 
\frac{d}{d\xi }\left( a^{2}\ln a^{2}\right) -v^{2}\xi ^{2}\frac{d}{d\xi}a^{2}\right] e^{-\xi ^{2}}d\xi,  \label{eq30}
\end{equation}%
where $\xi _{0}\rightarrow \infty $.

Further, using Eqs. (\ref{eq27}) and (\ref{eq28}), it is easy to verify that%
\begin{eqnarray}
{}&&v^{2}\frac{d^{2}a^{2}}{d\xi ^{2}}=\frac{d^{2}a^{2}}{d\tau ^{2}}=\frac{%
d^{2}a^{2}}{d\theta ^{2}}\theta _{\tau }^{2} \nonumber\\
{}&=&4V_{\max }-2a^{2}+4a^{2}\ln a^{2}+O(\varkappa ).  \label{eq31}
\end{eqnarray}
In what follows, we use this relation to exclude $a^{2}\ln a^{2}$ from
calculations. Thus, integrating in (\ref{eq30}) by parts and using (\ref%
{eq31}), we obtain%

\vspace{-8pt}

\begin{equation}
\zeta =-\frac{\varkappa }{4}e^{3-\rho ^{2}}\left[ v^{2}e^{-\xi ^{2}}\left( 
\frac{1}{2}\frac{d^{2}}{d\xi ^{2}}+\xi \frac{d}{d\xi }\right) a^{2}-\left(
1-v^{2}\right) \int_{\xi }^{\infty }\frac{da^{2}}{d\xi }e^{-\xi ^{2}}d\xi %
\right] +O(\varkappa ^{2}).  \label{eq32}
\end{equation}
Now we substitute $\psi $, $\chi $ and $\zeta $ into Eq. (\ref{eq18}), use
Eq. (\ref{eq31}) and integrate over $\xi $ by parts. This gives%
\begin{eqnarray}
&&\mkern -12mu \eta =\frac{\varkappa }{4v^{2}}e^{3-\beta ^{2}}{\Bigg \{}\sqrt{\pi }V_{\max }%
\left[ \left( 1+v^{2}\right) \frac{\xi \func{erf}\xi }{\rho ^{2}}-e^{\beta^{2}}\left( \left( 3v^{2}-1\right) \frac{\xi }{\rho ^{2}}\int_{0}^{\xi }%
\frac{\func{erf}\rho }{\rho }\,d\xi +\frac{\func{erf}\rho }{\rho }\right)\right] \nonumber\\
&&\mkern -12mu-v^{2}e^{-\xi ^{2}}\left( a^{2}+v^{2}\frac{\xi }{2\rho ^{2}}\frac{da^{2}}{d\xi }\right)
-\frac{1\!-\!v^{2}}{\rho ^{2}}{\Big [}\left( v^{2}\!-\!2\beta ^{2}\right) \xi
\int_{\xi }^{\infty }\!a^{2}e^{-\xi ^{2}}d\xi  \nonumber\\
&&\mkern -12mu-2\left[ \left( 1-v^{2}\right)
\xi ^{2}\!-\!\beta ^{2}\right] \!\int_{\xi }^{\infty }\!a^{2}\xi e^{-\xi ^{2}}d\xi\!+\!2\left( 1\!-\!v^{2}\right) \xi \!\int_{\xi }^{\infty }\!a^{2}\xi ^{2}e^{-\xi
^{2}}d\xi {\Big ]}\!+\!const\,\frac{\xi }{\rho ^{2}}{\Bigg \}}\!+\!O(\varkappa^{2}).  \label{eq33}
\end{eqnarray}%

\noindent Here, when calculating, we used the identity%
\begin{equation}
\beta ^{2}\int \frac{\,\mathrm{erf}\,\rho }{\rho }\frac{d\xi }{\xi ^{2}}%
=e^{-\beta ^{2}}\frac{2}{\sqrt{\pi }}\int e^{-\xi ^{2}}d\xi -\frac{\rho \,%
\mathrm{erf}\,\rho }{\xi }  \label{eq33a}
\end{equation}%
and replaced the indefinite integrals of regular functions by the definite
integrals over the interval $(0,\xi )$ plus constants.

We turn now to Eq. (\ref{eq20}). Using Eqs. (\ref{eq26}), (\ref{eq32}), (\ref%
{eq33}) and taking into account (\ref{eq31}), we find%
\begin{eqnarray}
&&2\chi -\zeta -2\eta =\varkappa e^{3-\beta ^{2}}{\Bigg \{}\frac{\sqrt{\pi }%
V_{\max }}{2v^{2}}{\Bigg [}\frac{\xi }{\rho ^{2}}\left( \left(
3v^{2}-1\right) e^{\beta ^{2}}\int_{0}^{\xi }\frac{\func{erf}\rho }{\rho }%
\,d\xi -\left( 1+v^{2}\right) \func{erf}\xi \right)\nonumber\\ 
&&+\left( 1-v^{2}\right) e^{\beta ^{2}}\frac{\func{erf}\rho }{\rho }{\Bigg ]}-%
\frac{v^{2}}{4}e^{-\xi ^{2}}\left[ \frac{1}{2}\frac{d^{2}a^{2}}{d\xi ^{2}}%
-\left( 1+\frac{1}{\rho ^{2}}\right) \xi \frac{da^{2}}{d\xi }\right] +\frac{1}{4}\left( 1-v^{2}\right) a^{2}e^{-\xi ^{2}}\nonumber\\
&&+\frac{1-v^{2}}{2v^{2}}\bigg[ \left( v^{2}-2\beta ^{2}\right) \frac{\xi }{\rho ^{2}}\int_{\xi }^{\infty }a^{2}e^{-\xi ^{2}}d\xi 
-\left( 2-v^{2}\right)\left( 1-\frac{2\beta ^{2}}{\rho ^{2}}\right) \int_{\xi }^{\infty }a^{2}\xi e^{-\xi ^{2}}d\xi \nonumber\\
&&+2\left( 1-v^{2}\right) \frac{\xi }{\rho ^{2}}\int_{\xi}^{\infty }a^{2}\xi ^{2}e^{-\xi ^{2}}d\xi \bigg]
+const\,\frac{\xi }{\rho ^{2}}{\Bigg \}}+O(\varkappa ^{2}).  \label{eq35}
\end{eqnarray}

\noindent Finally, we substitute this expression into Eq. (\ref{eq20}), integrate by parts
and use the identities%

\begin{equation}
\int_{-\infty }^{\infty }\frac{\xi }{\rho ^{4}}\left( \int_{0}^{\xi }\frac{\,
\mathrm{erf}\,\rho }{\rho }\,d\xi \right) \,d\xi = \frac{1}{\beta ^{2}}\left(1-e^{-\beta ^{2}}\right) 
+\frac{2}{\sqrt{\pi }}e^{-\beta ^{2}}\int_{0}^{\infty }%
\frac{e^{-\xi ^{2}}}{\xi ^{2}+\beta ^{2}}\,d\xi ,  \label{eq36}
\end{equation}%
\begin{equation}
\int_{-\infty }^{\infty }\frac{\xi \,\mathrm{erf}\,\xi }{\rho ^{4}}\,d\xi =%
\frac{2}{\sqrt{\pi }}\int_{0}^{\infty }\frac{e^{-\xi ^{2}}}{\xi ^{2}+\beta
^{2}}\,d\xi .  \label{eq37}
\end{equation}%

\noindent As a result, we obtain a simple formula for the deflection angle,%
\begin{eqnarray}
\Delta \varphi &=&\varkappa \frac{e^{3}\sqrt{\pi }V_{\max }}{2\beta v^{2}}%
\left( 1+v^{2}\right) \left( 1-e^{-\beta ^{2}}\right)
+\varkappa \frac{1-v^{2}}{2v^{2}}\beta \,e^{3-\beta ^{2}}\int_{-\infty
}^{\infty }a^{2}e^{-\xi ^{2}}d\xi +O(\varkappa ^{2})\quad\nonumber \\
&=&\frac{2GM}{bv^{2}}\left( 1+v^{2}\right) \left( 1-e^{-m^{2}b^{2}}\right)%
+2\pi G\sigma ^{2}{mb}\,e^{3-m^{2}b^{2}}\,\frac{1\!-\!v^{2}}{v^{2}}\!\int_{-\infty
}^{\infty }\!\!\!a^{2}e^{-\xi ^{2}}d\xi + O(\varkappa ^{2}),\quad  \label{eq38}
\end{eqnarray}%
where 
\begin{equation}
M=\left( e\sqrt{\pi }\right) ^{3}\sigma ^{2}m^{-1}V_{\max }\left(
1+O(\varkappa )\right)  \label{eq39}
\end{equation}%
is the total mass of the lump.

The first term in (\ref{eq38}) is the Schwarzschild deflection angle (\ref%
{eq23}) multiplied by the factor $1-e^{-m^{2}b^{2}}$ which takes into
account the mass distribution. Therefore, the resulting formula is valid for
any values of the impact parameter $b$. In particular, for $b=0$ we get 
$\Delta \varphi =0$, which is quite natural. However, regardless of the
values of $b$, the formula is valid only for sufficiently large initial
velocities such that $v^{2}\gg \varkappa V_{\max }=4\pi G\sigma ^{2}V_{\max} $. 
Otherwise, the deflection angle becomes significant, which contradicts our initial assumptions.

The second term in (\ref{eq38}) describes the periodic variations of the
deflection angle. In the integrand, the function $a(\theta )$ is found from
Eqs. (\ref{eq27}), (\ref{eq28}) followed by the substitution $\theta =\left(
1+\varkappa \Omega \right) (\tau _{R}-\xi /v)$. Therefore, after integration
over $\xi $, this term becomes a $T/2$-periodic function of $\theta
_{R}=\left( 1+\varkappa \Omega \right) \tau _{R}$. Note that this term becomes 
small for ultrarelativistic particles and vanishes when $v\rightarrow 1$, %
that is, the pulsations of the lump do not
affect the deflection of light. As emphasized in \cite{Koutvitsky4}, this
fact is a specific feature of the logarithmic potential (\ref{eq24}).

In the case of small oscillations, i.e., for $a_{\max }^{2}\ll 1$, we find $%
a(\theta )\approx a_{\max }\cos $ $\omega \theta $ with $\omega =2\pi
/T\approx \sqrt{1-\ln a_{\max }^{2}}$. Formula (\ref{eq38}) then gives
\begin{equation}
\Delta \varphi \approx \varkappa \frac{e^{3}\sqrt{\pi }}{4\beta v^{2}}%
a_{\max }^{2}\left[ \omega ^{2}\Big( 1\!+\!v^{2}\right) \left( 1\!-\!e^{-\beta^{2}}\right)
+\left( 1\!-\!v^{2}\right) \beta ^{2}e^{-\beta ^{2}}\left(1\!+\!e^{-(\omega /v)^{2}}\cos 2\omega \theta _{R}\right) \Big].\label{eq40}
\end{equation}

In general, averaging (\ref{eq38}) over the period, we find%
\begin{equation}
\overline{\Delta \varphi }=\varkappa \frac{e^{3}\sqrt{\pi }}{2\beta v^{2}}
\Big[ V_{\max }\left( 1+v^{2}\right) \left( 1-e^{-\beta ^{2}}\right)
+\left( 1-v^{2}\right) \beta ^{2}e^{-\beta ^{2}}\overline{a^{2}}\Big]+O(\varkappa ^{2}),  \label{eq41}
\end{equation}%
where $\overline{a^{2}}$, as well as $V_{\max }$, is a function of only $%
a_{\max }$. In Fig. 3 is shown the dependence of $\overline{\Delta \varphi }$
on the impact parameter for different values of $a_{\max }^{2}$. 

\begin{figure}[htb]
\includegraphics[width=8.5cm]{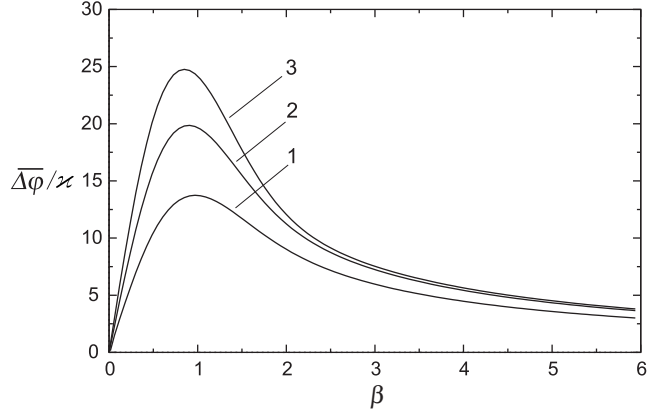}
\caption{Dependence of $\overline{\Delta \varphi }$ on the impact parameter for  
$a_{\max }^{2}=0.42$ (1), $a_{\max }^{2}=0.705$ (2), and $a_{\max }^{2}=0.86$ (3); $v=0.8$.}
\end{figure}
It can be
seen that at large $\beta $, the deflection angle behaves in the same way as
in the case of the Schwarzschild metric. Fig. 4 shows the deviation of the
deflection angle from its averaged value as a function of $\theta _{R}$. As
we can see, even with a sufficiently large $a_{\max }^{2}$ the oscillations
practically do not differ from sinusoidal ones.

\begin{figure}[htb]
\includegraphics[width=8.5cm]{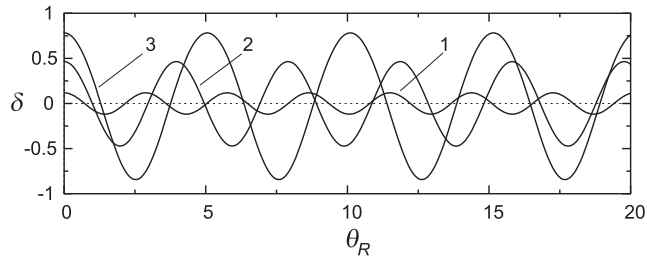}
\caption{Deviation of the deflection angle from its averaged value, $\delta=(\Delta \varphi-\overline{\Delta \varphi})/\varkappa$, for  
$a_{\max }^{2}=0.42$ (1), $a_{\max }^{2}=0.705$ (2), and $ a_{\max }^{2}=0.86$ (3); $v=0.8, \,\beta=1$.}
\end{figure}

\section{Discussion and concluding remarks}

Thus we have considered infinite motion of test particles in time-periodic
spherically symmetric spacetimes. Applying the perturbative approach to the
geodesic equations, we have obtained general formulas (\ref{eq17})-(\ref%
{eq19}) describing infinite trajectories of the particles passing through
oscillating dark matter. From these, formula (\ref{eq20}) immediately
follows for the total deflection angle of a particle coming from infinity,
passing through an oscillating distribution of matter and going to infinity.

As an example, we calculated the deflection angle of a particle passing
through an oscillating lump of the self-gravitating scalar field with the
logarithmic potential (\ref{eq24}). The result is given by Eqs. (\ref{eq38})-(\ref{eq41}).\ 
It should be noted that the stability of the scalar field lump we were
dealing with essentially depends on the amplitude of the oscillations. It
turned out that in certain narrow intervals of $a_{\max }$ values, the
solutions of the Einstein--Klein--Gordon system with high accuracy retain
their periodicity, making hundreds of oscillations, while outside them the
solutions, remaining well localized, lose their coherence \cite%
{Koutvitsky2}. This is also true without self-gravity effects \cite%
{Koutvitsky0, Koutvitsky5}. We assume that $a_{\max }$ belongs to one of
these intervals of quasistability. 

The distributions of scalar field dark matter (SFDM) we are considering can
be formed by axion-like particles in the ground state determined by the
dynamical balance of self-gravity, self-interaction, and quantum pressure.
The size of such a structure depends both on the mass of scalar particles
and on the self-interaction potential of the scalar field.

It was argued (see \cite{Hui} and references therein) that in studying DM structures on galactic
scales and above, the self-interaction of axion-like particles 
can be neglected. In this case, the central part of the dark matter
configuration, the so-called core, can be described by the system of Schr\"{o}dinger-Poisson
(SP) equations. The characteristic size of this core is roughly equal to the de Broglie
wavelength and amounts to $\sim$ 1 kpc for $m\simeq 10^{-22}$ eV, the core
mass being limited from above by the value of $10^{12}{M}_\odot$ \cite{Hui}.

On the scales larger than the de Broglie wavelength, the SFDM 
behaves as cold dark matter, and thus, the solitonic core should be surrounded 
by a scalar field halo with Navarro-Frenk-White (NFW) density distribution \cite{NFW} 
derived from the results of the N-particle modeling. Formation of the soliton-like core 
in the central region of the SFDM lump was clearly demonstrated in the 3D SP simulation 
of the ultralight dark matter \cite{Schive}, where a good fit was provided of the core 
density profile. These calculations, however, proved to be unable to yield the NFW density 
profile outside the core.

A comprehensive method for predicting the global density profiles of the
SFDM halo was proposed in \cite{Robles}. It enables to match the fit \cite{Schive} to the 
NFW profile. Comparison with circular velocities of the galaxies from the SPARC database
with those corresponding to this global density fit shows, however, that this new profile, while
provides better agreement with SPARC data at outer radii of galaxies, cannot solve and even exacerbates the central
density problem. This fact prompted the authors of \cite{Robles} to regard baryonic feedback 
as a probable candidate for resolving this discrepancy.

Another approach to the problem of the galactic core is based on the
assumption that the core consists of a central black hole surrounded by a
self-gravitating scalar field. For the case of a massive real scalar field
without self-interaction, oscillating long-lived self-gravitating
configurations of the scalar field around a non-rotating black hole were
found in Ref. \cite{Sanch} by
numerically solving the Einstein-Klein-Gordon equations. For a complex
non-self-interacting scalar field, such configurations were found in Ref.
\cite{Barranco3} in the form of self-gravitating coherent states.
The authors believe that these configurations can be detected due to their
influence on behavior of light rays, stars, gas clouds or other compact
objects surrounding a black hole in the center of galaxies. In this context,
we note that our results obtained in the weak field approximation lose their
validity near the horizon of the black hole, but remain valid for large
values of the impact parameter $b \gg r_g$, if the
oscillation amplitude of the scalar field is not too large.

The inclusion of self-interaction in the consideration can significantly
change the expected properties of dark matter distributions. Therefore, the
shape and parameters of the scalar field potential should be chosen from the
observational data. Thus, in Ref. \cite{Amaro} strong restrictions on
the axion mass and self-interaction coupling constant based on astrophysical
and cosmological observations were found for the $\phi ^{2}-\phi ^{4}$
potential. It turned out that these restrictions do not allow such a
potential to be suitable for describing both dark matter halos and compact
dark matter objects like boson stars. As was shown in Ref. \cite{Arbey}, the
mass of the scalar field and the self-interaction constant in this potential 
almost uniquely determine the characteristic scale of the scalar lumps. This means
that if these parameters are fixed, then all the halos considered in this
model as giant boson stars would have practically the same size, in stark disagreement
with observations. Apparently, the only way to overcome this difficulty is
to assume that galactic halos are collisionless ensembles of small-scale
components of scalar dark matter, rather than whole giant lumps of the
scalar dark matter field. These components, the so-called mini-massive compact halo objects \cite{Hernandez}, 
static or oscillating, should be of star-size or smaller
and have a mass less than $10^{-7}M_{\odot }$ following from microlensing
data \cite{Alcock}. This idea was further developed in Ref. \cite{Barranco4}, 
where the Einstein-Klein-Gordon system with the axion self-interaction potential 
was solved numerically in the quasi-classical limit. Having chosen the axion mass $m\sim 10^{-5}$ eV 
and the decay constant $f\sim 6\times 10^{20}$ eV, the authors found the self-gravitating
lumps of the axion field with the mass of an asteroid ($\sim 10^{-16}M_{\odot }$) and radius of a few meters.

In our toy model with logarithmic potential (24), it is assumed that the
self-interaction dominates gravity, since we work in the weak field
approximation. 
In the limiting case when gravity is neglected, the considered
oscillating lump becomes the exact pulson solution of the Klein-Gordon
equation with the characteristic size $\sim m^{-1}$. The inclusion of weak
gravity practically does not change the size of the lump, so that its
compactness remains the same, $\sim \left( e\sqrt{\pi }\right) ^{3}\sigma
^{2}V_{\max }$, where $\sigma^{2}\ll G^{-1}$ is assumed  (see \cite{Koutvitsky2} for details). Thus,
at $m\sim 10^{-22}$ eV, the lump size is about $0.06$ pc, that is much
smaller than the galactic core size, and much larger than the star size. As
for the period $T_{g}$, it can be seen from Fig. 2 that in a rather wide
range of $a_{\max }^{2}$values, we can take $T\simeq 10$, so that $T_{g}\approx \left( 2m\right) ^{-1}T\sim 1$ year. 
To obtain the Sun-sized lump we need to assume $m\simeq 2.7\times 10^{-16}$ eV, 
that gives $T_{g}\simeq 12$ seconds. Such lumps must have a very low average density 
$\rho \lesssim 10^{-7}\rho _{\odot }$, which means the weakness of the
gravitational field everywhere, including their interior. Nevertheless, we
believe that multiple gravitational scattering of particles by an ensemble
of these lumps with random oscillation phases can make an additional
contribution to the isotropization of cosmic rays and cause small variations
of neutrino flux from supernovae.

\section*{Acknowledgment}
We would like to thank the referee for useful comments and criticism, 
as well for drawing our attention to papers \cite{Barranco4, Amaro, Barranco3, Robles}.

\end{document}